\documentclass{article}

\usepackage{microtype}
\usepackage{graphicx}
\usepackage{subfigure}
\usepackage{booktabs} 

\usepackage{hyperref}

\usepackage[accepted]{icml2020}

\usepackage{times}
\usepackage{latexsym}
\RequirePackage{fix-cm}
\usepackage{mathptmx}      
\usepackage{relsize}
\usepackage{footnote}
\makesavenoteenv{tabular}
\makesavenoteenv{table}
\usepackage{amsmath, mathtools}
\usepackage{array, url}

\usepackage{dblfloatfix}
\graphicspath{{figures/}}
\DeclareGraphicsExtensions{.pdf,.jpg,.png}

\icmltitlerunning{}

\begin{document}

\twocolumn[
\icmltitle{Design Considerations for High Impact, Automated Echocardiogram Analysis}



\icmlsetsymbol{equal}{*}

\begin{icmlauthorlist}
\icmlauthor{Wiebke Toussaint}{equal,dssg}
\icmlauthor{Dave Van Veen}{equal,dssg}
\icmlauthor{Courtney Irwin}{equal,dssg}
\icmlauthor{Yoni Nachmany}{equal,dssg}
\icmlauthor{Manuel Barreiro-Perez}{husa,cibercv}
\icmlauthor{Elena D\'iaz-Pel\'aez}{husa,cibercv}
\icmlauthor{Sara Guerreiro de Sousa}{dssg}
\icmlauthor{Liliana Mill\'an}{dssg}
\icmlauthor{Pedro L. S\'anchez}{husa,cibercv}
\icmlauthor{Antonio S\'anchez-Puente}{cibercv,husa}
\icmlauthor{Jes\'us Sampedro-G\'omez}{cibercv,husa}
\icmlauthor{P. Ignacio Dorado-D\'iaz}{cibercv,husa}
\icmlauthor{V\'ictor Vicente-Palacios}{philips,husa}
\end{icmlauthorlist}

\icmlaffiliation{dssg}{Data Science for Social Good Fellowship 2019, Imperial College, London, UK}
\icmlaffiliation{husa}{Department of Cardiology, Hospital Universitario de Salamanca-IBSAL, Universidad de Salamanca, Spain}
\icmlaffiliation{cibercv}{CIBERCV, Instituto de Salud Carlos III, Madrid, Spain}
\icmlaffiliation{philips}{Philips Healthcare}

\icmlcorrespondingauthor{Wiebke Toussaint}{w.toussaint@tudelft.nl}

\icmlkeywords{}

\vskip 0.3in
]



\printAffiliationsAndNotice{\icmlEqualContribution} 

\begin{abstract}

Deep learning has the potential to automate echocardiogram analysis for early detection of heart disease. Based on a qualitative analysis of design concerns, this study suggests that 
predicting normal heart function instead of disease accounts for data quality bias and significantly increases efficiency in cardiologists' workflows.

\end{abstract}

\section{Introduction}

Cardiovascular diseases take the lives of 17.9 million people every year, accounting for 31\% of all global deaths \cite{who2017cvd}. Many consequences of heart disease could be mitigated with early diagnosis. Echocardiograms are ultrasound video scans of the heart that are non-invasive, side-effect free, cheaper and faster to perform than other imaging techniques. They are the most frequently used cardiac imaging test for the screening, diagnosis, management, and follow-up of patients with suspected or known heart diseases. The left ventricle ejection fraction (LVEF) measures the percentage of blood leaving the heart each time it contracts. It can be calculated from echocardiograms and is used as a prognostic indicator to identify and follow the progression of heart disease \cite{Drazner2011}. 

Automating the analysis of echocardiogram scans is considered an important task for increasing access to health care \cite{loh2017deep}, particularly for early disease detection in remote and low resource areas \cite{zhang2018fully}. Further benefits lie in improving the efficiency of medical care and for establishing effective public health programmes \cite{MeleroAlegriae2019salmanticor}. However, the motivation for the use of deep learning in echocardiogram analysis is typically the potential to reduce human error and outperform human experts. Recent research has indeed shown that convolutional neural network models can surpass human-level performance for estimating the LVEF \cite{ouyang2020video}. \citet{bizopoulos2019survey} found that studies using deep learning for echocardiography tended to be highly variable in terms of the research questions that were asked, and inconsistent in the metrics used to report results. 

The goal of this study was to perform a qualitative analysis of the impact of automating echocardiogram analysis at the Department of Cardiology at the Hospital Universitario de Salamanca-IBSAL (HUSA) in Northern Spain. A secondary objective was to build a pipeline for automated, end-to-end echocardiogram analysis with deep learning for HUSA. This paper discusses the design process and discoveries, their impact on global health and on defining the prediction goal for the pipeline, and the results achieved.
\section{Design Process}

Human-centred design offers a practical approach to arriving at innovative design solutions. Methods from IDEO.org's Design Kit \cite{ideo} were used as practical tools to analyze the potential impact of automating echocardiogram analysis with deep learning. At first, five guiding questions were formulated during a brainstorming session:

\begin{enumerate}
\setlength\itemsep{-0.2em}
    \item What is the context for echocardiography at HUSA?    
    \item How is an echocardiogram study performed?
    \item What is the role of the cardiologist in this process?
    \item How are study results used for decision making?
    \item Who is impacted by automating the analysis process?
\end{enumerate}

Stakeholder mapping, testing assumptions with \textit{``5 Whys''}, mapping the journey of a cardiologist and performing a conceptual analysis of deep learning for echocardiogram analysis within a clinical setting were selected as appropriate tools for arriving at answers to the guiding questions. Qualitative information to answer these questions was gathered over a series of video calls and a one-week site visit to HUSA, which included participation in a routine clinic visit to outlying rural areas, observing the process of taking echocardiogram scans of healthy, recovering and sick patients, and participating in specialists' morning meetings where upcoming medical procedures and decisions are discussed. 
\section{Design Discoveries}

\paragraph{Context}

The cardiology imaging unit at HUSA serves over 40 patients per day and performs over 10,000 echocardiogram studies every year. 40\% of the patients examined at HUSA have normal heart function. Cardiologists also perform echocardiogram studies during weekly visits to rural Spanish health centers to reduce the public transit burden of patients that come for routine checkup or on referral from their physician. 80\% of these patients require no further intervention. If the study indicates abnormal heart function, patients are referred to HUSA for further screenings.

\paragraph{Echocardiogram Procedure}

During a study, the cardiologist captures different views of the heart from specific vantage points for several heartbeats on video. A typical study takes 10 -- 20 minutes. After performing the study, the cardiologist analyses the video scans to calculate measurements that indicate heart condition. To calculate LVEF, the cardiologist selects the appropriate scan, identifies image frames corresponding to one cardiac cycle's end-systole and end-diastole, and manually traces the heart chamber in both frames. A software programme subsequently calculates the LVEF. The process is repeated and averaged over 5 heartbeats. A skilled specialist can annotate a study in 8 -- 10 minutes, cumulating to 4 -- 5 hours of annotations per day.

\paragraph{Decision Making}

At HUSA, a LVEF of less than 40\% is considered abnormal, 40\% to 60\% grey zone, and above 60\% normal. For patients with a normal LVEF and no medical preconditions that put them at risk, deviation in the LVEF is tolerable. Measurement deviations become a concern when they are in the grey zone, especially when the patient's medical history gives reason to suspect heart disease. Patients at risk of heart disease will receive regular checkups, while serious cases will go for follow-up assessments with more expensive equipment. Treatment decisions are typically made by a multidisciplinary team of experts who consider the health condition of the patient holistically.

\paragraph{Data Quality and Variability}

Echocardiogram studies are guided by a standardized medical protocol. Nonetheless, variability exists in studies, making automated analysis more challenging than for other imaging modalities. The specialist's skill affects the scan quality. Occasionally a scan is retaken due to quality issues, resulting in multiple scans for the same view. The anatomy and health status of the patient further affect data quality. The scans of older and unhealthy patients, who are at greater risk of heart disease, are frequently of poorer quality due to rib protrusions, obesity or movement during the procedure. Image quality also varies across scanner models. HUSA employs multiple different models simultaneously. Outlying clinics have older models, which produce lower quality scans.

\paragraph{Implications of Discoveries}

Predictive results are adversely affected by poor data quality. Variability in data quality biases predictive accuracy against patients that are more inclined to have a heart condition and that cannot access newer scanner models. Leveraging this insight, by using automated analysis to filter out patients with normal heart function, cardiologists can better allocate their time and expertise. This could lead to positive downstream effects, such as focusing more attention on sick patients. In addition, this could enable the public health system to increase its capacity for routine checkups for early diagnosis.

\section{Deep Learning Pipeline and Results}

HUSA's echocardiogram dataset consists of 25,000 annotated studies collected over 10 years. Each study contains video data in the DICOM format. Segmentation masks and measurements were manually recorded by cardiologists and have been used to derive labels. Anonymized patient metadata is available in a supplementary database. Data cleaning, pre-processing and filtering was done to ensure consistent data representation and sampling.

The deep learning pipeline \footnote{\url{https://github.com/dssg/usal_echo_public}} replicates a cardiologist's process of analyzing an echogardiogram in three steps: (1)~\textit{classify views}: parasternal long-axis, apical 2-chamber, and apical 4-chamber; (2)~\textit{segment chambers}: left ventricle and left atrium; (3)~\textit{estimate measurement}: LVEF. The pipeline deploys the models of \citet{zhang2018fully}, using a VGGNet \cite{simonyan2014very} for classification and U-net \cite{ronneberger2015u} for segmentation. The LVEF is then calculated from the segmentation masks. 

Based on the design discoveries, the objective of the deep learning pipeline is to identify patients with normal heart function. Thus, when executed in sequence, the pipeline predicts heart health to be normal, grey zone or abnormal. Classification results had an accuracy of 98\%. The segmentation model DICE scores ranged from $0.83$ -- $0.88$ depending on view and heart chamber. After estimating the LVEF, the pipeline predicts normal heart condition with 80\% precision and 30\% sensitivity. High precision is necessary to ensure that patients at risk receive care. Sensitivity indicates the potential time saving for cardiologists.

\section{Conclusion}

Deep learning has the potential for automating echocardiogram analysis, the most frequently used imaging technique for early detection of heart disease. While existing studies focus on matching or surpassing experts' ability to predict disease, this paper suggests that predicting normal heart function instead aligns data quality with the prediction objective and significantly reduces cardiologists' time investment in patients that do not need their expertise. This can pave the way for large-scale, low-cost preventative heart screenings, while reducing the time burden on skilled experts.

\section*{Acknowledgements}
This project was supported by the Department of Cardiology at the Hospital Universitario de Salamanca-IBSAL, the Data Science for Social Good Foundation, and the Gandhi Centre for Inclusive Innovation at Imperial College, London. 

\bibliography{library.bib}
\bibliographystyle{icml2020}

\end{document}